\documentclass[epsfig,showpacs,floatfix]{revtex4}
\usepackage{epsfig,colordvi}

\usepackage{epsfig}
\usepackage{graphicx}
\usepackage{graphics}

 \def\be{\begin{equation}}
 \def\ee{\end{equation}}
 \def\bea{\begin{eqnarray}}
 \def\eea{\end{eqnarray}}
 \def\bean{\begin{eqnarray*}}
 \def\eean{\end{eqnarray*}}

\begin{document}
\title{Core - Corona Model describes the Centrality Dependence of $v_2/\epsilon$. }
\author{J. Aichelin and K. Werner}
\affiliation{SUBATECH, Laboratoire de Physique Subatomique et des
Technologies Associ\'ees, \\
Universit\'e de Nantes - IN2P3/CNRS - Ecole des Mines de Nantes \\
4 rue Alfred Kastler, F-44072 Nantes, Cedex 03, France\\}
\begin{abstract}
Event by event EPOS calculations in which the expansion of the system is described by
{\it ideal} hydrodynamics reproduce well the measured centrality dependence of $v_2/\epsilon_{part}$,
although it has been claimed that only viscous hydrodynamics can reproduce these data.
This is due to the core - corona effect which manifests itself in the initial condition
of the hydrodynamical expansion. The centrality dependence of $v_2/\epsilon_{part}$
can be understood in the recently advanced core-corona model, a simple parameter free
EPOS inspired model to describe the centrality dependence of different observables from
SPS to RHIC energies. This model has already been successfully applied
to understand the centrality dependence of multiplicities and of the average transverse
momentum of identified particles.

\end{abstract}
\date{\today} \maketitle

Ever since Drescher et al. \cite{Drescher:2007cd} have claimed that the elliptic flow,$ \frac{v_2}{\epsilon}$, observed in the RHIC experiments, is
even for the most central collisions at least 25\% below the ideal hydrodynamical limit, the centrality dependence of
the elliptic flow is considered as one of the key variables for the understanding of ultrarelativistic heavy ion collisions.
Before this finding one had concluded from other observables that the Quark Gluon Plasma, created in the energetic reaction,
is a perfect liquid \cite{Muller:2007rs}.

\begin{figure}
\begin{center}
\hspace*{-0.5cm}
\includegraphics[width=6.5cm]{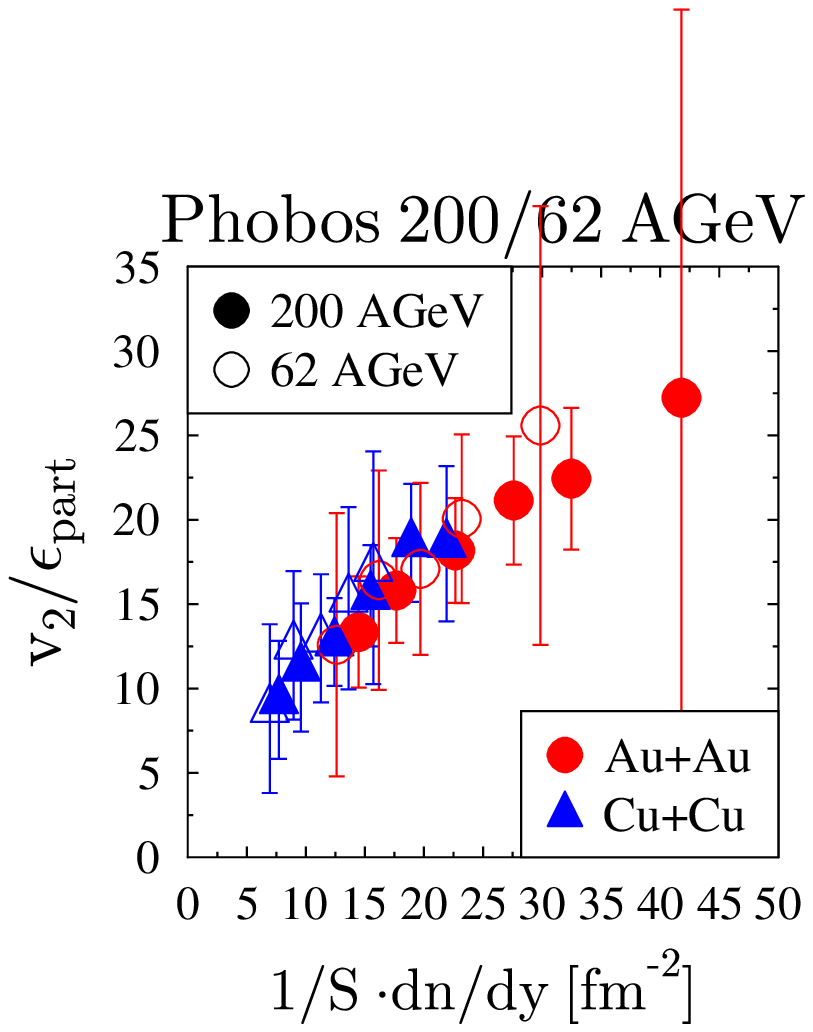}
\includegraphics[width=6.5cm]{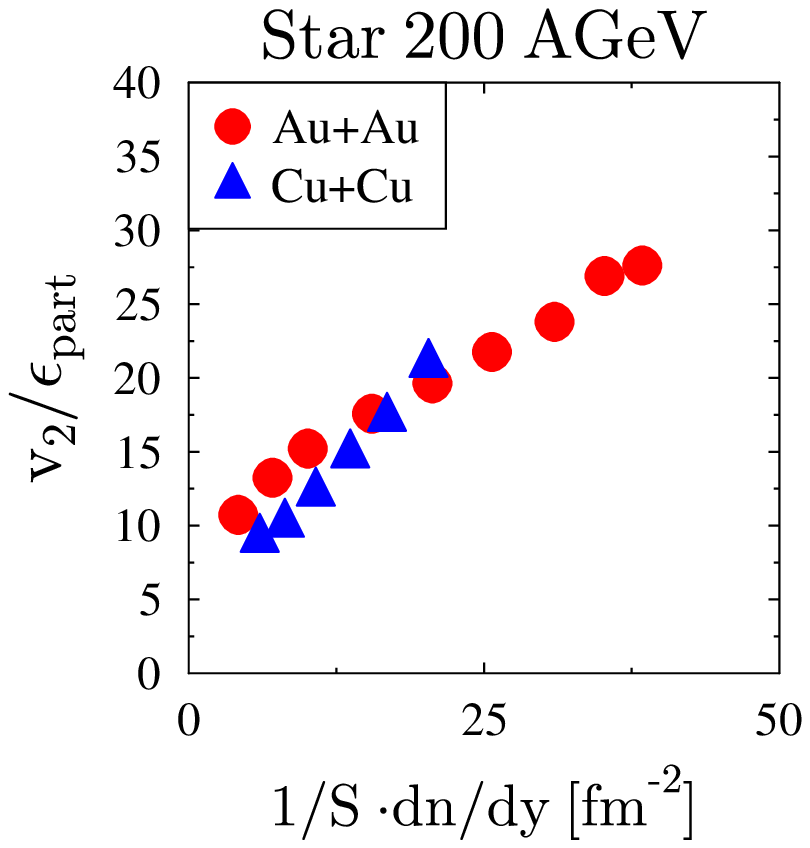}
\end{center}
\caption{Dependence of $v_2/\epsilon_{part}$ on  $\frac{1}{S}\frac{dn}{dy}$ in AuAu and CuCu collisions at 200 AGeV in comparison with the results of the STAR experiment \cite{starpr,star:2008ed} and for 200 AGeV/62 AGeV in comparison with the  Phobos data \cite{Alver:2006wh}. } \label{v2ef}
\end{figure}

The analysis in \cite{Drescher:2007cd} is based on the assumption that the centrality dependence of the elliptic flow
can be described by a simple formula \cite{Bhalerao:2005mm}:
\be
\frac{v_2}{\epsilon} = \frac{v_2^{hydro}}{\epsilon}\frac{1}{1+\frac{K}{K_0}}.
\ee
 $v_2$ is the measured elliptic flow, $v_2 = <cos2(\phi- \phi_R)>$, where $\phi$ ($\phi_R $) is the azimuthal angle of the emitted particle (reaction plane). $v_2^{hydro}$, the value of the elliptic flow expected in the hydrodynamical limit, and $K_0$ are
 the two free parameters of the model. The average is performed over many events which belong to the same centrality class of the reaction. $\epsilon$ is the spatial participant eccentricity of the particles in an event at the beginning of the hydrodynamical expansion, defined as $
\epsilon=\frac{\sqrt{\left(\sigma_y^2-\sigma_x^2\right)^2+4\sigma^2_{xy}}}{\sigma^2_y+\sigma^2_x}$
with $\sigma^2_{x}=<x^2> -<x>^2$,
$\sigma^2_{y}=<y^2> -<y>^2$,
$\sigma_{xy}=<x y>-<x><y>$.
It has to be calculated in a theoretical model. K is the Knudsen number defined as
\be
\frac{1}{K} = \frac{\sigma}{S}\frac{dN}{dy}c_s.
\label{knud1}
\ee
with $\sigma$ being the partonic cross section, $c_s$ being the sound velocity and S being the transverse area =
$4 \pi \sqrt{\sigma_x^2 \sigma_y^2-\sigma^2_{xy}}$. Using a Glauber model to determine S and $\epsilon$, the measured \cite{Alver:2008ck} (phobos) or calculated values \cite{Ray:2007av} (star) values of $dn/d\eta$ as well as the experimentally
observed $v_2$ values of the Phobos \cite{Alver:2006wh}, the Star \cite{starpr,star:2008ed} and the Phenix
collaboration  \cite{phenix,Shimomura:2009hb} one can calculate $\frac{v_2}{\epsilon}$ as a function of $dn/d\eta$.
The ratio between $dn/d\eta$ and $dn/dy$ depends on the particle species but is not far from 1. We assume here  $dn/d\eta = dn/dy$
because our conclusions do not depend on the precise value of this ratio.
The result of this analysis is shown in fig. \ref{v2ef}.

On the first view it seems that the data of different colliding systems at different energies fall in this representation on top of each other and therefore the Knudsen number depends indeed only on $\frac{1}{S}\frac{dN}{dy}$. However, if one regards the left figure in more detail one realizes that at a given centrality $ \frac{v_2}{\epsilon}$ is almost independent of the beam energy \cite{Alver:2006wh}. $dn/d\eta$ for 200 AGeV is about 1.4 times larger than that observed for 62 AGeV, almost independent of the centrality \cite{Alver:2008ck}. Thus the Knudsen number extracted from the 200 AGeV data is 40\% smaller than that from the
62 AGeV data. This 40\% change of the Knudsen number is hardly visible in the plot which means that this presentation is not very sensitive to the Knudsen number.
That for the same value of $dn/d\eta$ the value of $\frac{v_2}{\epsilon}$ of the 62 AGeV data is systematically higher than that of the 200 AGeV data can be attributed to different sound velocities \cite{Bhalerao:2005mm}, however for the price that an additional free parameter is employed. The speed of sound depends on the density. The density profile is different at different energies but approximately the same for different systems at different energies. It remains to be explained why at the lower beam energy the system is closer to the hydrodynamical limit. However, the error bars are too large to allow for firm conclusions. The scaling of $\frac{v_2}{\epsilon}$ with $dn/d\eta$ is questioned by the more recent Star data, displayed on the right hand side, where only statistical errors are shown. The Star collaboration has analyzed
also peripheral AuAu events, in contradistinction to the Phobos collaboration. Therefore these data allow a more detailed
comparison of both systems. We see for both systems an almost linear dependence of  $\frac{v_2}{\epsilon}$ with $dn/d\eta$. The slope for the CuCu system is, however, clearly different from that for AuAu, what one could have guessed already from the Phobos data \cite{Drescher:2007cd}.

By fitting the experimental results with the theoretical curve, eq. \ref{knud1}, one can determine $K/(\sigma c_s)$ and
finally the Knudsen number K of each centrality bin,
assuming (centrality independent) values for $\sigma$ and $c_s$.
Even for the most central bins $\frac{v_2}{\epsilon}$ depends on $dn/d\eta$. Thus
the hydrodynamical limit K=0 is not reached. From the extrapolation towards $dn/d\eta \to \infty$, corresponding to $K\to 0$, the authors concluded that the observed flow is at least 25 \% below the limit for ideal hydrodynamics.

These findings triggered the development of codes using
viscous hydrodynamics. Luzum et al.\cite{Luzum:2008cw}
showed in viscous hydrodynamical calculations that different viscosities yield a different centrality dependence of $v_2/\epsilon_{part}$ and of $<p_T>$. This dependence they used to identify that viscosity value which describes the 200 AGeV AuAu data. More recently, Song et al. \cite{Song:2008si} have extended the calculation to the CuCu system.

What has not been realized in this line of arguments is the importance of the initial configuration at the moment when the
hydrodynamical expansion starts. The equations of ideal hydrodynamics predict how
- for a given equation of state - this initial configuration develops  in time under the assumption that the system is locally in equilibrium. Therefore it is expected that different initial conditions give different final distributions. The initial configuration at the beginning of the hydrodynamical expansion is unknown. It is difficult to asses because the time scale for equilibration in the initial state is much faster than predicted by microscopical calculations.

Recently one realized that event by event fluctuations of the initial conditions are not only much stronger than
initially expected but also that they manifest themselves in the observables \cite{Andrade:2009em,Hama:2009vu}. Only the particles deep inside the interaction region, the core particles, collide sufficiently frequent to form a locally equilibrated source whereas those close to the surface, called corona particles, do not come to a local equilibrium and do therefore not take part in the hydrodynamical evolution of the system.

EPOS is a consistent quantum mechanical multiple scattering approach
based on partons and strings~\cite{nexus}, where cross sections
and the particle production are calculated consistently, taking into
account energy conservation in both cases (unlike other models where
energy conservation is not considered for cross section
calculations~\cite{hladik}).
A special feature is the explicit treatment of projectile and target
remnants, leading to a very good description of baryon and antibaryon
production as measured in proton-proton collisions at 158~GeV at
CERN~\cite{nex-bar}. Nuclear effects related to CRONIN transverse
momentum broadening, parton saturation, and screening have been introduced
into EPOS~\cite{splitting}.

In heavy ion collisions
(and more recently also in proton-proton collisions)
collective behavior is taken into
account~\cite{corona}, in the following fashion:
the initial scatterings, as described above,
lead to the formation of strings, which break into segments,
usually identified with hadrons. When it comes to heavy ion collisions,
the procedure is modified: one considers the situation at an early
proper time $\tau_{0}$, long before the hadrons are formed: one
distinguishes
between string segments in dense areas (more than some critical density
$\rho_{0}$ segments per unit volume), from those in low density areas.
The high density areas are referred to as core, the low density areas
as corona \cite{corona}. It is important to note that initial conditions
from EPOS are based on strings, providing a ``flux-tube'' like
structure in case of individual events (a single flux tube in case
of many overlayed events). Based on the four-momenta of the string
segments which constitute the core, we compute the energy density
$\varepsilon(\tau_{0},\vec{x})$ and the flow velocity
$\vec{v}(\tau_{0},\vec{x})$.

\begin{figure}[ht]
\begin{center}
\hspace*{-0.5cm}
\includegraphics[width=9.5cm,angle=-90]{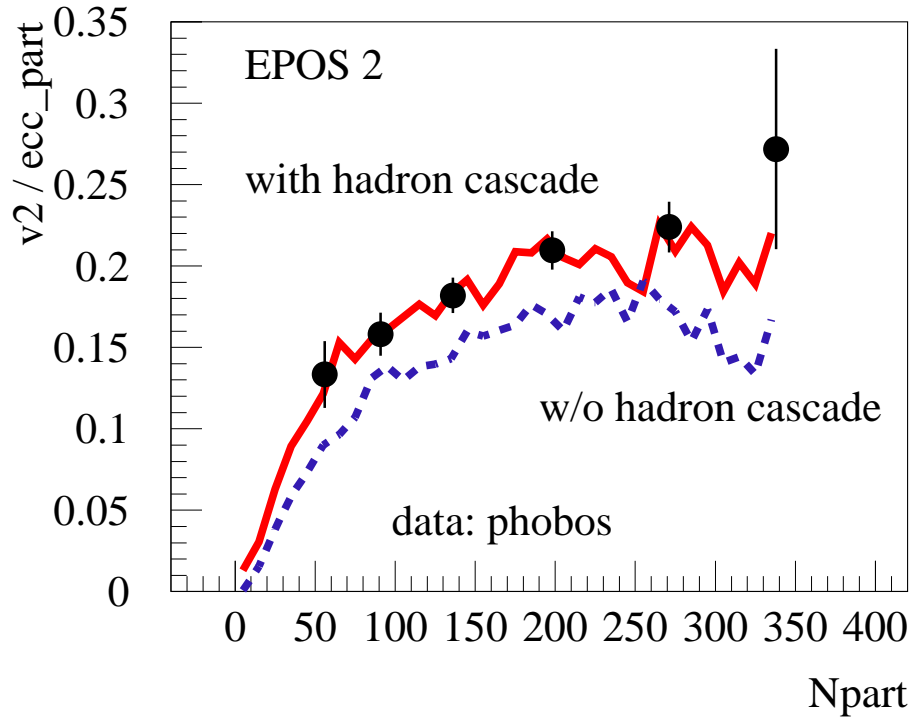}
\end{center}
\caption{$v_2/\epsilon_{part}$ for AuAu collsions at $\sqrt{s}=200 GeV$, as predicted by EPOS, compared with
data. The curves are the result of event - by event hydrodynamical calculations with a freeze out temperature
of 166 MeV with (solid) and without (dashed) hadronic rescattering.} \label{ex88}
\end{figure}

Having fixed the initial conditions, the system evolves according to
the equations of {\it ideal} hydrodynamics. Hadronization occurs finally
based on the Cooper-Frye prescription. The centrality dependence of $ \frac{v_2}{\epsilon}$
of the EPOS calculation for AuAu at 200 AGeV is compared with the experimental results in fig. \ref{ex88}. We see an almost perfect agreement despite of the fact that EPOS uses ideal hydrodynamics. Especially the absolute value of $v_2^{hydro}/epsilon$, the free
parameter in the core corona model, is well
reproduced. The centrality dependence is due to the core-corona effect
in the initial configuration \cite{corona}, which manifests itself also in other observables.

Recently we have developed a simple model \cite{Aichelin:2008mi,Aichelin:2010ed} to study the consequences of this core-corona effect and
to interpret the very involved EPOS simulations. In this model we define corona particles as those nucleons which have only one initial collision whereas the others are considered as core
particles. $f_{core}$ is the fraction of core nucleons which depends on the centrality, the system size and (weakly) on the beam energy. This fraction is determined in the same Glauber model which is used to calculate the eccentricity $\epsilon$ and the transverse area S. Core particles form a locally equilibrated source whereas corona nucleons are treated like elementary
proton-proton collisions with no interactions with their environment. Of course this is a very crude model but it has the
advantage that no free parameter is needed in this superposition of pp collisions and a thermalized source. The
present experimental error bars do leave little room for an improvement as has been shown in \cite{Aichelin:2008mi,Aichelin:2010ed}.

In this simple model we could show that, independent of the system size, the centrality dependence of the multiplicity
$M^i$ and of the $<p_T^i>$ of all identified hadrons from SPS to RHIC energies can quantitatively be described  by
\begin{eqnarray}
 & &M^i(N_{\rm part})\label{eq1}\\
 & & = N_{\rm part}\,\big[f_{core}\cdot M^i_{\rm core}+(1-f_{core})\cdot M^i_{\rm corona}\big]
\,\nonumber\\
& &<p_T^i>(N_{\rm part})\label{eq2}\\
& & = N_{\rm part}\,\big[f_{core}\cdot <p^i_{T\ \rm core}>+(1-f_{core})\cdot <p^i_{T\ \rm corona}>\big].
\,\nonumber
\end{eqnarray}
For our calculation, we fix $M_{\rm core}\ (<p^i_{T\ \rm core}>)$ by applying eq. \ref{eq1} (\ref{eq2})to the
most central AuAu or PbPb data point. $M_{\rm corona}$ and $<p^i_{T\ \rm corona}>$ are
given as half of the multiplicity and the $<p_T^i>$ measured in pp collisions. Once these
parameters are fixed, the centrality dependence of $M^i$ and $<p_T^i>$ is determined by eq. \ref{eq1} and \ref{eq2}.
Especially the centrality dependence of the observables for the lighter CuCu system follows then
without any further input.
\begin{figure}[ht]
\begin{center}
\hspace*{-0.5cm}
\includegraphics[width=9.5cm]{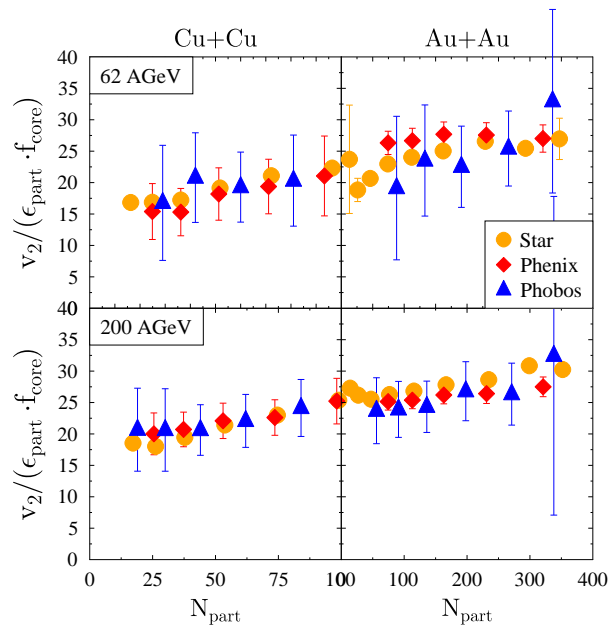}
\end{center}
\caption{$v_2/\epsilon_{part}(N_{part})/ f_{core}(N_{part})$ as a function of $N_{part}$ for the data at 62 (top) and 200 AGeV
(bottom). The data are from refs.  \cite{starpr,star:2008ed,vol} (Star) , \cite{Alver:2006wh} (Phobos) and \cite{phenix,Shimomura:2009hb} (Phenix).}
\label{ex99}
\end{figure}
Both, the core - corona model as well as  the approach of Drescher et al. \cite{Drescher:2007cd}, assume that
the system is not in the hydrodynamical limit, however the reason is different. In the approach of \cite{Drescher:2007cd} it is assumed that all particles come to the same degree of thermalization independent of whether they are
at the surface or in the interior of the interaction zone and that this thermalization is not complete.
The degree of thermalization depends on the centrality. In the core - corona
model the particles in the interior of the reaction zone have sufficient collisions to reach complete thermalization whereas
those located at the surface do not came to equilibrium at all. They behave like being created in independent
NN collisions. In this separation it follows the findings of the EPOS program. The centrality dependence is due to the centrality dependence of $f_{core}$.

In the core - corona model only the core particles feel the eccentricity of the overlap region. The particles
produced by corona nucleons show an isotropic distribution in azimuthal direction.
Therefore, in the core - corona model, $v_2/\epsilon_{part}$ is expected to be $\propto f_{core}$
\be
v_2/\epsilon_{part}(N_{part}) = f_{core}(N_{part}) \frac{v_2^{hydro}}{\epsilon}.
\ee
Because $f_{core}(N_{part})$ and $\epsilon_{part}$ is calculated in the Glauber model $v_2^{hydro}$ is the only free parameter.
The core - corona model allows for some immediate predictions. Because $f_{core}$ is similar in central CuCu and AuAu collisions (Fig. \ref{ex99}) we expect that central collisions of AuAu and CuCu show a similar $v_2/\epsilon_{part}$ despite of the large difference of $N_{part}$. Therefore $v_2/\epsilon_{part}$ plotted as a function of $\frac{1}{S}\frac{dn}{dy}$ is in central  CuCu
data larger as compared to AuAu data. In peripheral CuCu and AuAu collision $v_2/\epsilon_{part}$ should be very similar for the same $N_{part}$.
The easiest way to compare our results with experiments is to divide the experimental results by $f_{core}(N_{part})$.
In this presentation we divided out the centrality dependence and therefore we expect that the data are horizontal straight lines.

In Fig. \ref{ex99} all presently available experimental results are presented in that way.
We see that in this presentation the 200 AGeV data of Phenix and Phobos, presented in the bottom part, follow a straight line.
The Star data show a small slope. We should stress that for the peripheral CuCu data where the core contains about 15 nucleons
we cannot expect that our assumptions are fully justified and one may doubt whether there the core forms really a fully thermalized source. Also the data at 62 AGeV, shown in the top panels, are compatible with our prediction that $v_2^{hydro}/\epsilon_{part}= v_2/\epsilon_{part}(N_{part})/ f_{core}$ is constant. However, whereas the Phobos data are compatible
with the assumptions that $v_2^{hydro}/\epsilon_{part}$ is the same for AuAu and CuCu, the Phenix data show a difference of 30\%
between CuCu and AuAu. This is not understood yet.

\begin{figure}[ht]
\begin{center}
\hspace*{-0.5cm}
\includegraphics[width=9.5cm]{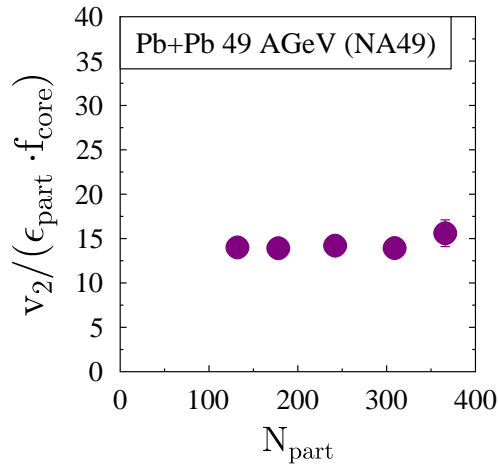}
\end{center}
\caption{$v_2/\epsilon_{part}(N_{part})/ f_{core}(N_{part})$ as a function of $N_{part}$ for the NA49 data \cite{Alt:2003ab} at $\sqrt{s}=17.2 GeV$.}
\label{v2ena}
\end{figure}

Our model can be extended to the lower SPS energies where $v_2/\epsilon_{part}(N_{part})$ has been measured by the
NA49 collaboration \cite{Alt:2003ab}. These results, show in fig. \ref{v2ena}, are also compatible with a straight line.

In conclusions, we have shown that the experimentally observed centrality dependence of $v_2/\epsilon_{part}(N_{part})$ is well described by the EPOS event generator using {\it ideal} hydrodynamics. In contradistinction to other hydrodynamical calculations
EPOS creates its initial condition which fluctuates from event by event. Only those regions of the interaction zone which have
a density above a threshold density form a locally equilibrated source whereas the others are treated as independent NN collisions.
The simple EPOS inspired core - corona model, which has successfully described the centrality dependence of the multiplicity
and the average transverse momentum of identified particles, can also quantitatively explain the centrality dependence of $v_2/\epsilon_{part}(N_{part})$. Only core particles come to a local thermal equilibrium and develop a $v_2$ whereas the corona
particles decay isotropically. The fraction of both classes of particles depends on the centrality and the centrality dependence of the elliptic flow reflects directly this dependence. The only free parameter is $v_2^{hydro}$. 
This results questions the idea to use the centrality dependence of $v_2$ to determine the viscosity of a quark gluon plasma.
First of all, on the conceptual level,  it is difficult to imagine that particles close the surface of the interaction zone behave exactly as those in in the center. Second, on the quantitative level, two additional free parameters, $K_0$ and $c_s(E)$, have to be introduced to relate the centrality dependence of $v_2$ with the expectations for a viscous system. The numerical value of both is difficult to asses by other observables.

 All approaches agree on the fact that the system is not a perfect liquid, even in central collisions. 
 In the core corona model this is due to the fact that nucleons at the surface of the interaction zone behave like pp collisions whereas the center comes to complete equilibrium. In the viscous hydrodynamical approach it is assumed that the interaction zone has no surface but that the whole system is described by a finite viscosity which varies with centrality.

Acknowledgements: We would like to thank Drs. R. Bellwied, U. Heinz, B. Jacak, J.Y. Ollitrault, S. Shi, R. Snellings,  A. Timmings
for valuable discussions and Dr.Voloshin to make available the preliminary AuAu data at 64 AGeV.

\end{document}